# Increasing trends in the severity of Australian fire weather conditions over the past century

*Soubhik Biswas*[A,B*], *Andrew J Dowdy*[C,D] *and Savin S Chand*[A,B,E,F]

[A]Institute of Innovation, Science and Sustainability, Federation University, Australia.

[B]Centre for New Energy Transition Research, Federation University, Australia.

[C]University of Melbourne, Melbourne, Victoria, Australia.

[D]ARC Centre of Excellence for 21st Century Weather, Melbourne, Victoria, Australia.

[E]Future Regions Research Centre, Federation University, Australia.

[F]Centre for Smart Analytics, Federation University, Australia.

[*]Corresponding author. Email: soubhik.biswas.pro@gmail.com

**Abstract:** Understanding how weather and climate influence fire risk is important for many purposes, including climate adaptation planning and decision-making in sectors such as emergency management, finance, health and infrastructure (e.g., for energy and water availability). In this study, bias-corrected 20CRv2c reanalysis data are used to investigate the climatology and long-term trends of weather conditions associated with landscape fires in Australia. The McArthur Forest Fire Danger Index (FFDI) is used here as a broad-scale representation of weather conditions known to influence fire behaviour based on wind speed, humidity, temperature and rainfall measures. In particular, using this reanalysis dataset allows analysis over a longer time period than previous studies, from 1876 to 2011. Another novel aspect is that trends are examined using several different approaches, including a method to help account for the influence of interannual drivers of climate variability not previously used for fire weather analysis. Results show increases in mean and extreme seasonal FFDI values throughout Australia in general, with all statistically significant trends being positive in sign for individual climate zones. Humidity and temperature trends, attributable to human-caused climate change, are shown to be the main cause of the increase in dangerous weather conditions for fires. These findings build on previous studies, with the novel data and methods used adding confidence to the overall understanding of fire risk factors in a changing climate.

**TOC Summary:** In the wake of increasing bushfire impacts in recent decades across Australia, it is important to understand the role played by human-caused climate change as distinct from natural variations in weather and climate conditions. This paper contributes towards this understanding by quantifying the increasing trends in the severity of fire weather conditions across Australia using a dataset longer than previous studies, as well as novel methods to account for natural variability. The findings are intended to help with disaster preparedness and management, as well as long-term plans and policies for climate adaptation.

# 1   Introduction

Australia is prone to extreme fires in the landscape (referred to as bushfires in Australia), which often pose a serious threat to people (Blanchi et al., 2014; Cameron et al., 2009; Victoria, 2017), properties (Blanchi et al., 2010), agriculture (Filkov et al., 2020) and the ecosystem (Canadell et al., 2021; Dickman & McDonald, 2020). From the perspective of prediction, disaster risk management and reduction, response and recovery, understanding climatological evolution of forest fires has become imperative.

One of the key factors that influences forest fire behaviour is the daily meteorological conditions, often referred to as the fire weather conditions (Bradstock, 2010). Internationally, several indices have been developed to quantify the fire weather condition using meteorological data, including the McArthur Forest Fire Danger Index (McArthur, 1967) (FFDI), the Canadian Forest Fire Weather Index System (Wagner, 1987) (FWI) and the United States National Fire Danger Ratings System (Bradshaw et al., 1983) (NFDRS). The FFDI is used for our study as a commonly understood index in Australia that provides a generalised combination of several weather conditions known to influence fire behaviour (based on wind speed, humidity, temperature and rainfall). While the FFDI has been used in many previous studies, here we use a novel dataset including FFDI data available back to 1876 based on 20CR reanalysis, as documented in previous studies (Biswas et al., 2025; Biswas et al., 2022).

Fire weather indices, such as the FFDI used here, are a useful way to combine several meteorological factors known to influence fire behaviour, while noting that there are many other factors than weather conditions that can contribute to fire occurrence and resultant impacts (including vegetation-related factors such as fuel type, as well as human-related factors such as ignition and suppression). Fire weather indices have been correlated with fire occurrence and impacts in previous studies (Blanchi et al., 2010; Canadell et al., 2021; Christoff, 2023; Goss et al., 2020), while noting that the focus of this particular study is on fire weather conditions as represented by FFDI, rather than specifically considering fire occurrence and impacts.

Previous studies have considered recent extreme cases as well as long-term trends in fire weather. For example, case studies examining the 2019-2020 Black Summer fires in eastern and southern Australia noted anomalous hydroclimatic variables and geomorphic characteristics contributing to the severity of those recent fires (Deb et al., 2020). Other studies of changes over longer time periods have been done using station observations (Deb et al., 2020) and gridded analyses of observations (Goss et al., 2020) through Australia, while noting that such studies do not extend back prior to about 1950 for available data. A key aim of this study is to extend the analysis of long-term trends back further than previous studies, using a dataset of FFDI that was produced based on the 20CR reanalysis (Alexander & Arblaster, 2009).

With a changing climate, including the observed increase in temperature over Australia (Alexander & Arblaster, 2009) it is anticipated that fire weather conditions are also likely to change (Canadell et al., 2021). For example, studies using observation-based data have shown that there have been significant increases in FFDI in Australia in recent decades (Dowdy, 2018; Harris et al., 2017). Moreover, various climate model outputs also project a very high likelihood of more dangerous fire weather conditions in the future (Dowdy et al., 2019; Jones et al., 2022) under increased greenhouse warming. Climate drivers such as El Niño Southern Oscillation (ENSO), Indian Ocean Dipole (IOD) and Southern Annular Mode (SAM) also have a significant influence on the Australian fire weather variability on a year-to-year time scale (Abram et al., 2021; Dowdy, 2018; Harris & Lucas, 2019).

Given that Australia is prone to extreme wildland-fire conditions, it is important to investigate the influence of climate change on fire weather conditions over this region. The results presented here are derived using bias-corrected and spatially interpolated Twentieth Century Reanalysis version 2c (20CRv2c) (Biswas et al., 2022) that provides long–term data of the input meteorological variables for FFDI from 1851-2014. As such, it enables the examination of fire weather climatology and trends over a longer time period than in previous studies (Abram et al., 2021; Clarke et al., 2013; Dowdy, 2018; Harris & Lucas, 2019).

## 2 Data and Methodology

The input variables for the computation of the FFDI are obtained from the 20CRv2c dataset (Compo et al., 2011), which is a daily and 2.0°×2.0° longitude-latitude gridded product for the period 1851 to 2014 available across 56 different ensemble members. Each variable from 13 randomly selected ensemble members was bilinearly interpolated to a 0.05 X 0.05-degree grid to be consistent with the observation-based reference dataset (Dowdy, 2018; Jones et al., 2009). Following the methodology detailed in previous research (Biswas et al., 2022), the interpolated data was then bias-corrected with respect to the observation-based input variables (i.e., rainfall, wind speed, temperature and relative humidity) using a ranking-based quantile-quantile matching approach (Biswas et al., 2022). Monotone piecewise cubic Hermite spline was used as a probability density function (PDF) to calibrate the spatially interpolated 20CRv2c data due to its better performance (Biswas et al., 2022).

Daily values of FFDI (McArthur, 1967; Noble et al., 1980) are computed on the interpolated and bias-corrected 20CRv2c data ranging from 1851 to 2014. The FFDI is based on a dimensionless drought factor $df$ (Biswas et al., 2022), wind speed $ws$ (km h$^{-1}$), relative humidity $rh$ (%) and temperature $t_{max}$ (°C) on a given day such that:

$$FFDI = \exp(0.0338 \times t_{max} - 0.0345 \times rh + 0.0234 \times ws + 0.243147) \times df^{0.987} \qquad (1)$$

The drought factor ($df$) is based on the soil moisture deficit and calculated here using the Keetch-Byram drought index (Keetch & Byram, 1968) using the past 20 days' rainfall (Finkele et al., 2006). We note that other fire weather indices were also considered as potential candidates for this analysis, including those with more complexity, such as the Canadian Fire Weather Index (FWI) System, which includes multiple moisture deficit metrics to represent fuel aridity (Wagner, 1987). However, the FFDI was used here, given it is most commonly used for climate analysis purposes in Australia, relating to its history of being designed using Australian conditions (McArthur, 1967) and then subsequently being used for many decades for operational fire management purposes. An additional key reason why FFDI is used in this study is that a FWI dataset based on 20CR reanalysis is not available as far as the authors are aware, whereas the FFDI dataset based on 20CR reanalysis used here had already been produced, as documented in previous studies (Biswas et al., 2025; Biswas et al., 2022). We also note that there are some general similarities in how FWI and FFDI represent fire danger, such as both indices having extremely high quantile values for severe fire events (Dowdy et al., 2009).

The study period used here is from 1876 -2011, which spans the available periods of data for the climate indices (i.e., the Niño 3.4 index as a metric for ENSO, the DMI index as a metric for IOD and the AOI index as a metric for SAM) (PSL, 2022; Smith, 2018, 2022) as well as the dataset of FFDI based on 20CR reanalysis, previously published (Biswas et al., 2025; Biswas et al., 2022), which extends back to 1876. We acknowledge the possibility to extend this research in potential future studies including with more recent years of data, as well as potential to consider climate modelling of projected future change, while noting that a key focus of this particular study is to consider FFDI trends back further in time than previous studies (i.e., back to 1876 in contrast to previous analyses in the literature). To remove any further inconsistencies, we used the ensemble mean values from a representative sample (previously discussed, randomly selected 13 ensemble members) of the calculated FFDI data to have better estimations of the certainty of results.

The following steps were then implemented to determine long-term trends in the Australian fire weather conditions. First, the daily FFDI data were grouped into four respective seasons: spring (September to November, SON), summer (December to February, DJF), winter (March to May, MAM) and autumn (June to August, JJA). National gridded seasonal mean values of FFDI are then calculated from the daily data for each year at each grid cell location (hereafter referred to as the seasonal mean FFDI). Next, we used the 'peak-over-threshold' approach to determine the number of days each season exceeds the $90^{th}$ percentile FFDI values (hereafter referred to as the seasonal number of extreme FFDI days). Other thresholds could also have been used in addition to the 90th percentile, while noting that the 90th percentile helps provide a reasonably large sample size as compared to if more extreme thresholds had been selected for this analysis, with the $90^{th}$ percentile being used in other climate applications (IPCC, 2023), for considering some extreme values.

In addition to examining long-term trends in the seasonal mean values and in the seasonal number of extreme FFDI days, using data as described above, we also examined these long-term trends using a modified version of the FFDI values. The modified FFDI values are designed to account for interannual variations associated with climate modes (such as ENSO, etc.), which is included here as an alternative analysis approach to help provide potential further insight on trends (e.g., similar to previous studies using a related approach to this[33]). The modified FFDI values were produced for each individual grid cell location as follows. First, we computed the slope coefficients (i.e., as an output from multivariate linear regression) between mean seasonal FFDI values and the climatic indices (i.e., NINO 3.4, DMI and AOI) (Bamston et al., 1997; BoM, 2012; Gong & Wang, 1999; Lim et al., 2021; Saji et al., 1999). We then calculated the residual FFDI values, $FFDI_{res}$, as shown in Equation 2, based on subtracting the regressed components for each of the climatic index values:

$$FFDI_{res} = FFDI - m_1 \times NINO - m_2 \times DMI - m_3 \times AOI \qquad (2)$$

where

$FFDI$ = Mean seasonal FFDI time-series

$FFDI_{res}$ = Residuals of seasonal FFDI time-series

$NINO$ = NINO 3.4 (Bamston et al., 1997) time-series (ENSO)

$DMI$ = DMI (BoM, 2012; Saji et al., 1999) time-series (IOD)

$AOI$ = SAM (Gong & Wang, 1999; Lim et al., 2021) time-series (SAM)

$m_1$ = Slope coefficient associated with $NINO$

$m_2$ = Slope coefficient associated with $DMI$

$m_3$ = Slope coefficient associated with $AOI$

Monthly values of all the climate indices (i.e., NINO 3.4, DMI and AOI) (PSL, 2022; Smith, 2018, 2022) for the above-mentioned climatic drivers are available between 1876 and 2011. The statistical significance of linear trends, after removing the effects of natural variability, was calculated at the 95% significance level using the Wald test (D. Ward & S. Ahlquist, 2018). To further validate our results from trend analysis, we have calculated the mean-state changes in both seasonal FFDI values and extreme FFDI days between two climatic epochs (somewhat arbitrarily selected): 1876-1945 and 1945-2011, where the latter period was selected here to approximately represent a period with somewhat more rapid industrialisation and anthropogenic climate change than the earlier period (IPCC, 2023).

For ease of interpretation, area-averaged trends – as well as the mean-state changes – are calculated for separate Natural Resource Management (NRM) regions (Whetton et al., 2015) (Fig 1), which largely corresponds to the broadscale climate and biophysical regions of Australia.

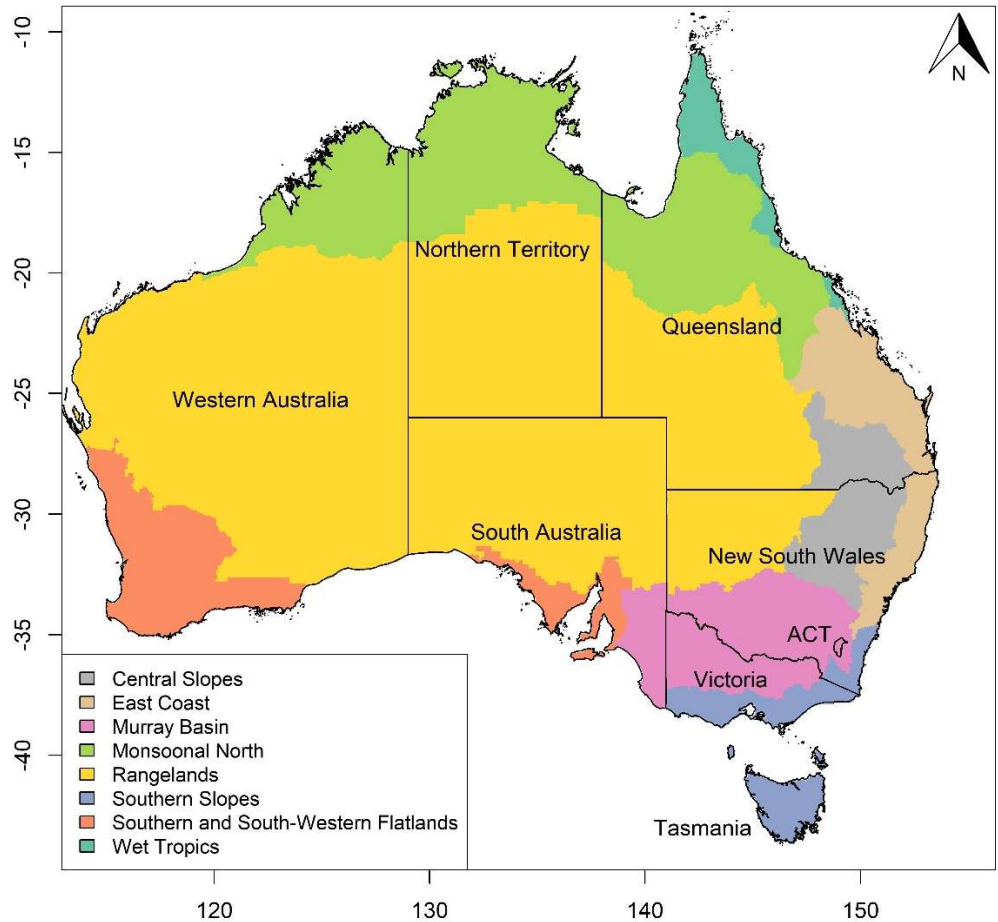

**Fig. 1.** Map of Australia showing different states, territories and National Resource Management (NRM) clusters (Whetton et al., 2015).

# 3 Results

*3.1 Climatological characteristics*

The climatological mean and the extreme FFDI days (i.e., using the $90^{th}$ percentile for this analysis) over the study period 1876 – 2011, computed at each individual grid point across Australia, are shown in Fig. 2. As expected, clear differences in the FFDI mean-state conditions are noted across seasons, and it is found to be consistent with past studies (Dowdy, 2018; Harris & Lucas, 2019; Luke & McArthur, 1978)

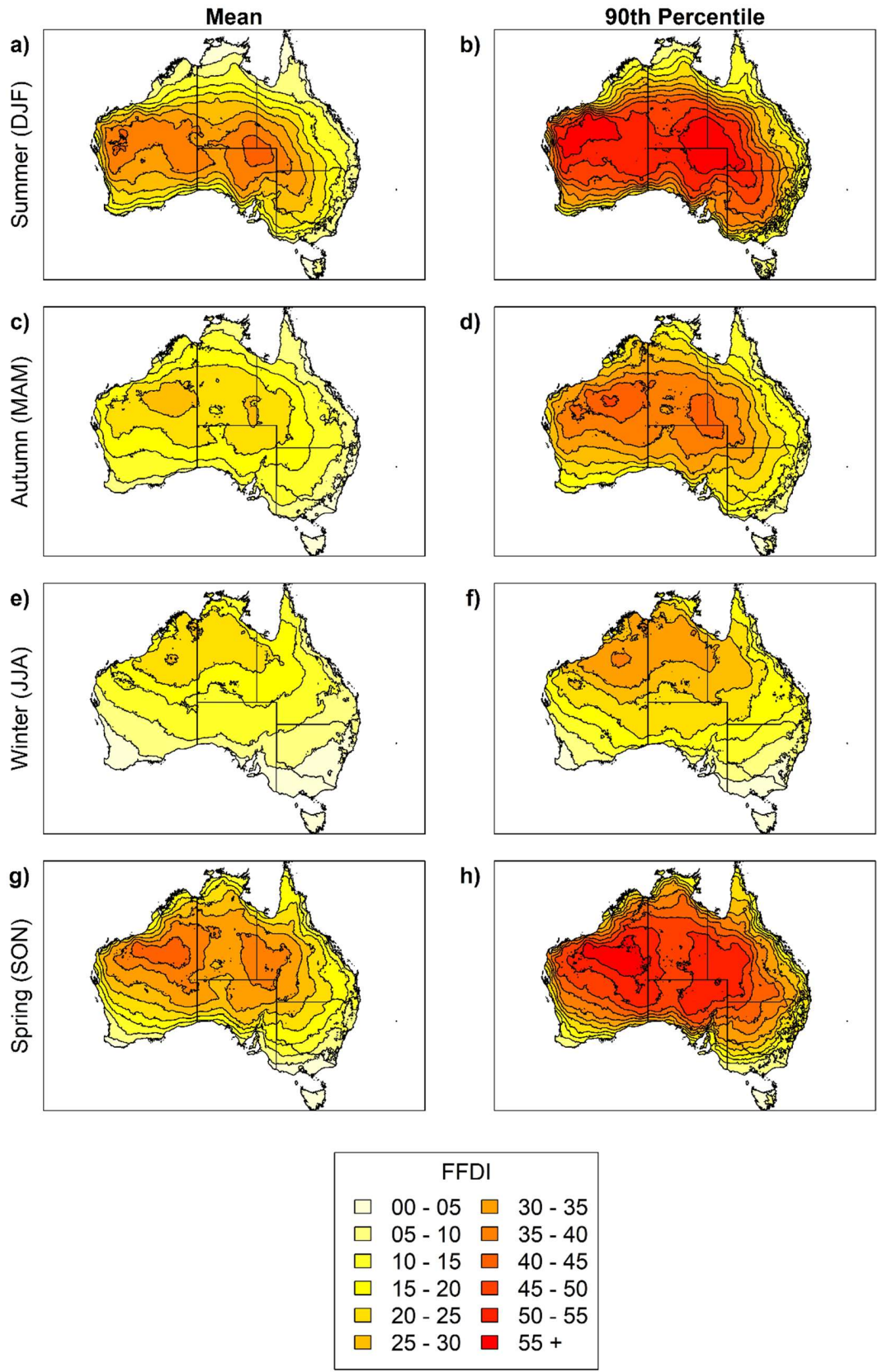

**Fig. 2.** Seasonal climatological mean and extreme (i.e., the 90th percentile) FFDI values across Australia over the period 1876 -2011.

Overall, our results show that the austral spring and summer seasons are more prone to elevated mean and extreme fire weather conditions across Australia, noting that relatively higher FFDI values can also be reported even during the autumn and winter seasons as these correspond with the tropical Australian dry season(e.g., parts of Western Australia, Northern Territory and Queensland). In addition, Australia's east coast was observed to be slightly more prone to elevated FFDI during the austral spring season than in the summer (Dowdy, 2018; Luke & McArthur, 1978).

*3.2    Long-term trends*

Long-term trends in the mean FFDI values and the number of extreme FFDI days for each season over the period 1876-2011 (after removing the influence of natural modes of climate variability, see Methods) are shown in Fig. 3 and 4.

Generally, statistically significant upward trends in the mean FFDI values are observed across Australia during each of the four seasons, except for the southern slopes and Murray basin during the austral spring, where a slight downward (but statistically insignificant) trend is noted (Fig. 3). In particular, a considerably strong statistically significant increasing trend was seen across the NRM Rangelands (> ~0.02 year$^{-1}$) and East Coast (> ~0.01) (see also Table 1). While trends are positive throughout Australia during the austral summer season, substantially strong upward trends are present mainly east of Western Australia's divide (Fig. 3a). Interestingly, the upward trends for the austral autumn and winter seasons are statistically significant for the entire Australian region (Figs. 3b, c). In the spring season, statistically significant upward trends are primarily noted in Western Australia and South Australia, as well as around the East Coast region near the Queensland and New South Wales (NSW) border (Fig. 3d). A key result shown in Table 1 is that all of the trends that are significant are positive in sign, with no negative trends being statistically significant at the 95% confidence level used throughout this study.

Results from the mean-state changes in seasonal FFDI values (Supplementary Fig. 1) between the two climatic epochs, 1877-1945 and 1945-2011, are generally consistent with the long-term trends (Fig. 3) for most parts of Australia (Table 1). For example, we see a statistically significant increase in the mean FFDI values for all seasons in the latter epoch, with exceptions for regions across eastern Australia (Supplementary Fig. 1d), where we note slightly suppressed (but statistically insignificant) FFDI conditions. Both methods show that FFDI values have increased over the recent period compared with the historical past.

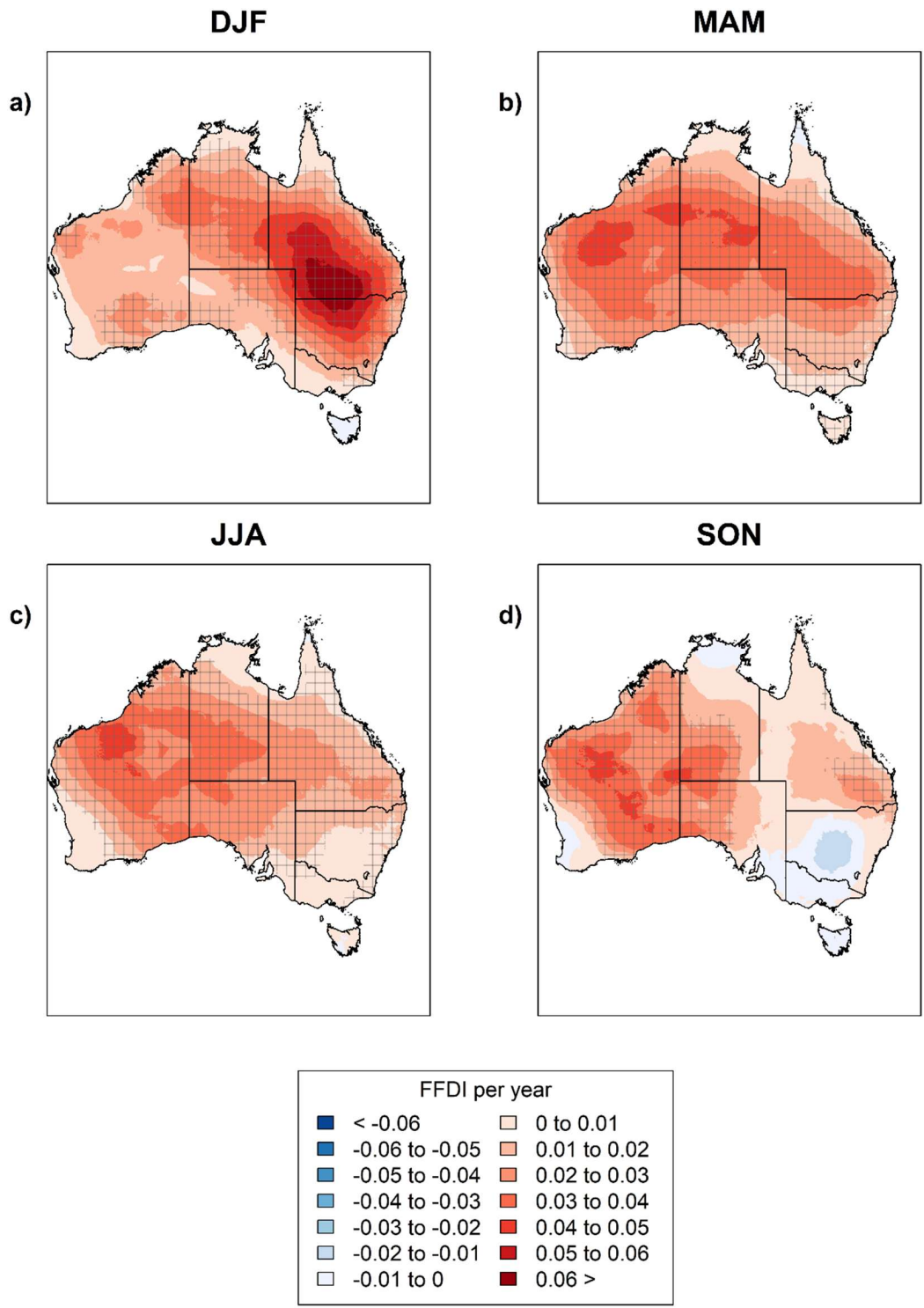

**Fig. 3.** Mean seasonal FFDI trend at each grid point for the study period 1876 – 2011. Statistically significant regions at 95% confidence level are shown as overlapping grey grids.

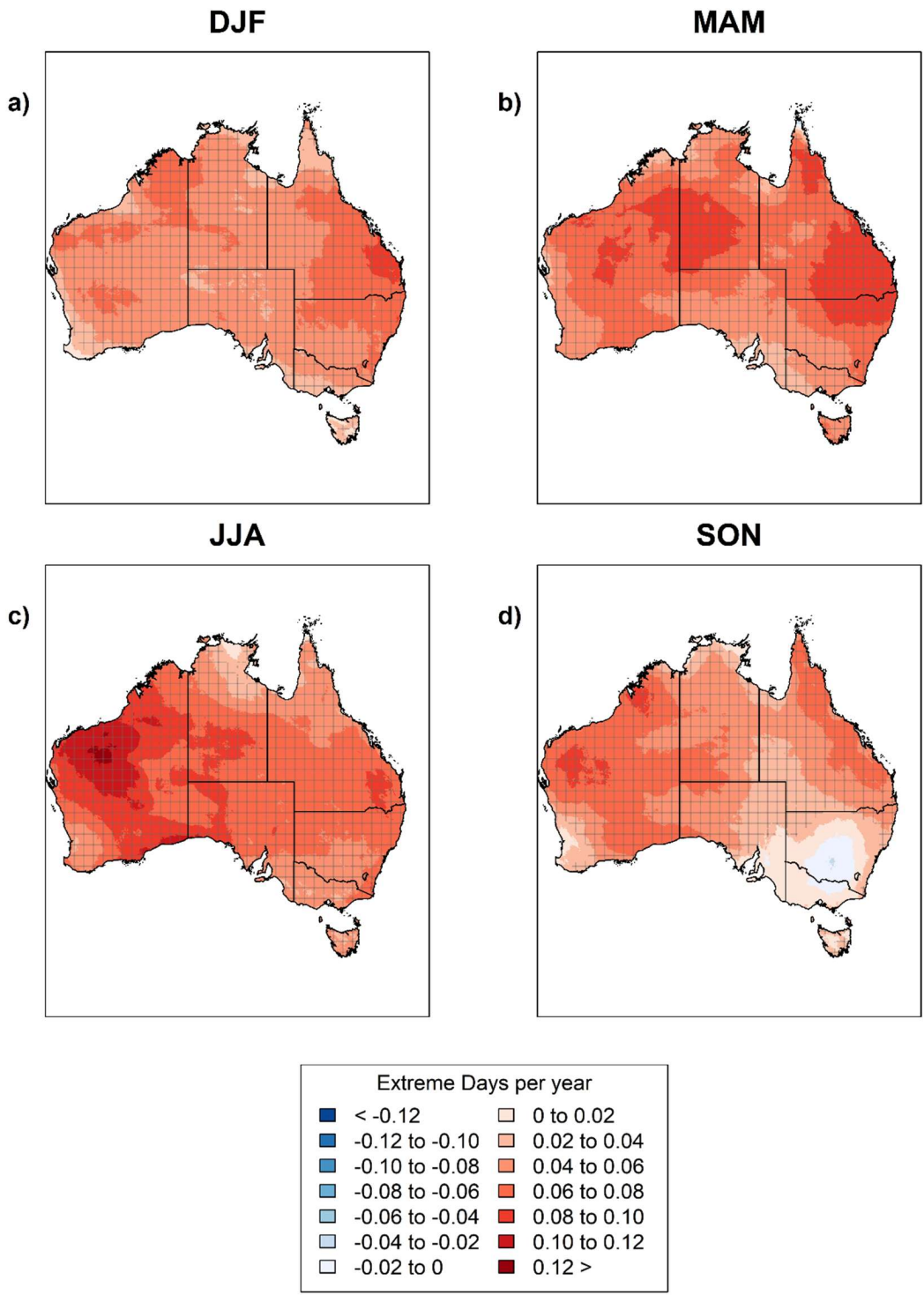

**Fig. 4.** As in Fig. 3 but for seasonal 90th percentile peak over threshold values of FFDI.

**Table 1.** Mean seasonal FFDI trend across each NRM region for the study period 1876 – 2011. The percentage change in seasonal mean FFDI values between 1876 and 1945 and 1945 and 2011 is shown in brackets. Statistically significant values at 95% confidence level are shown in bold.

|  | DJF | MAM | JJA | SON |
|---|---|---|---|---|
| Australia | **0.0245 (7.72%)** | **0.0251 (11.04%)** | **0.0195 (9.55%)** | **0.0155** (3.30%) |
| Central Slopes | **0.0448 (19.10%)** | **0.0264 (16.56%)** | **0.0135 (12.67%)** | 0.0083 (-3.25%) |
| East Coast | **0.0270 (19.31%)** | **0.0181 (14.34%)** | **0.0133 (11.38%)** | **0.0111** (1.45%) |
| Murray Basin | **0.0190 (6.50%)** | **0.0118 (7.40%)** | **0.0046 (8.55%)** | -0.0052 (-5.20%) |
| Monsoonal North | **0.0192 (15.17%)** | **0.0162** (7.74%) | **0.0142** (3.55%) | 0.0097 (1.15%) |
| Rangelands | **0.0276 (6.76%)** | **0.0318 (11.50%)** | **0.0260 (11.66%)** | **0.0220** (4.37%) |
| Southern Slopes | 0.0039 **(8.41%)** | 0.0049 **(11.42%)** | 0.0015 **(6.97%)** | -0.0001 (-0.63%) |
| Southern and South-Western Flatlands | 0.0082 **(4.76%)** | **0.0148 (13.45%)** | **0.0064 (14.03%)** | 0.0063 **(7.27%)** |
| Wet Tropics | 0.0044 (6.82%) | 0.0003 (-4.98%) | 0.0020 (0.52%) | 0.0035 (1.53%) |

These changes can be partially attributed to the trend in mean seasonal relative humidity values (Supplementary Fig. 3), while rainfall and temperature (Supplementary Fig. 4 and 6) were found to be the next most influencing input variables for trends in mean seasonal FFDI (Fig. 3). During the austral spring, we see a decline of at least 0.01% in relative humidity per year on average across Western Australia and the East Coast; while in summer, a statistically significant decreasing trend, no less than 0.02% in relative humidity per year, was observed over the Monsoonal North, Central Slopes and East Coast (Supplementary Fig. 3). The downward trends in mean seasonal FFDI (Fig. 3) across southeast Australia during austral spring coincide with decreasing mean seasonal temperature trends (Supplementary Fig. 6d) and increasing relative humidity (Supplementary Fig. 3d) and rainfall (Supplementary Fig. 4d) trends. For completeness, we have shown the mean seasonal values of wind speed (Supplementary Fig. 7) even though it was found not to influence the trends in mean seasonal FFDI much (Fig. 3).

Additionally, we did a time-series analysis on Lockhart (146.72° E, 35.22° S) to explore the reasons behind the downward trend (not statistically significant) for mean seasonal FFDI in southeast Australia during the austral spring season (Fig. 5). A time series for mean seasonal FFDI was constructed based on bias-corrected 20CRv2c18 from 1877 to 2011, and for FFDI using observation-based variables (Dowdy, 2018) from 1953 to 2011. To verify whether this downward trend is due to the removal of natural variability, we again repeated the time-series analysis without removing natural climatic variabilities based on the approach described by Equation 2, which was found not to significantly modify the long-term trend in FFDI (Fig. 5). A downward trend (~ -0.01 year$^{-1}$) still occurs for 20CRv2c FFDI for the entire time period (1877 – 2011), but when a trend is computed for the observation-based FFDI from 1953 to 2011, an increasing trend (~ 0.01 year$^{-1}$) was observed (Fig. 5a). Thus, it can be concluded that the long-term downward trends in mean seasonal FFDI are masking the more recent upward trends in the southern slopes and Murray basin. A downward trend (~ - 0.02 year$^{-1}$) for the entire time period and an increasing trend (~ 0.005 year$^{-1}$) for the observation-based temperature further validate that the long-term trends are indeed masking the more recent trends in southeast Australia (Fig. 5b). There could be another potential factor here; since we are using the FFDI dataset from 1876-2011, with more recent years not included in the study (Prelgauskas, 2016; Wilkinson et al., 2016), as scope for further potential research. Fig. 5a also shows that the larger FFDI values appear more frequently in recent decades, suggesting a potential acceleration in the occurrence frequency of elevated FFDI periods rather than it being a gradual linear change over this long time period examined using 20CR reanalysis here (Table 1). This is associated with a trend towards increased variability in more recent decades compared to earlier decades. For example, the standard deviation of FFDI values in the first epoch is 2.97 compared to 4.64 in the second epoch. Increased variability is also found for other regions of Australia and other seasons, such as the Murray Basin, Monsoonal North, Southern Slopes, Southern and South-Western Flatlands, in both summer and spring.

It is also noted that rainfall in Australia in increasing in variability in general, associated with more periods of drought as well as more of extreme rain (IPCC, 2023), noting examples in this study indicative of this (e.g., for rainfall in Fig. 5). This can make trend analysis challenging, with variations in trends depending on which metric is considered (e.g., percentiles or other averaging approaches). As shown in (Dowdy, 2018), there has been a trend towards increased variability in FFDI values for Australia, with the results shown here consistent with that, including noting increased rainfall variability (as a factor adding to variations in KBDI and DF used in the FFDI formulation: Eqn 1).

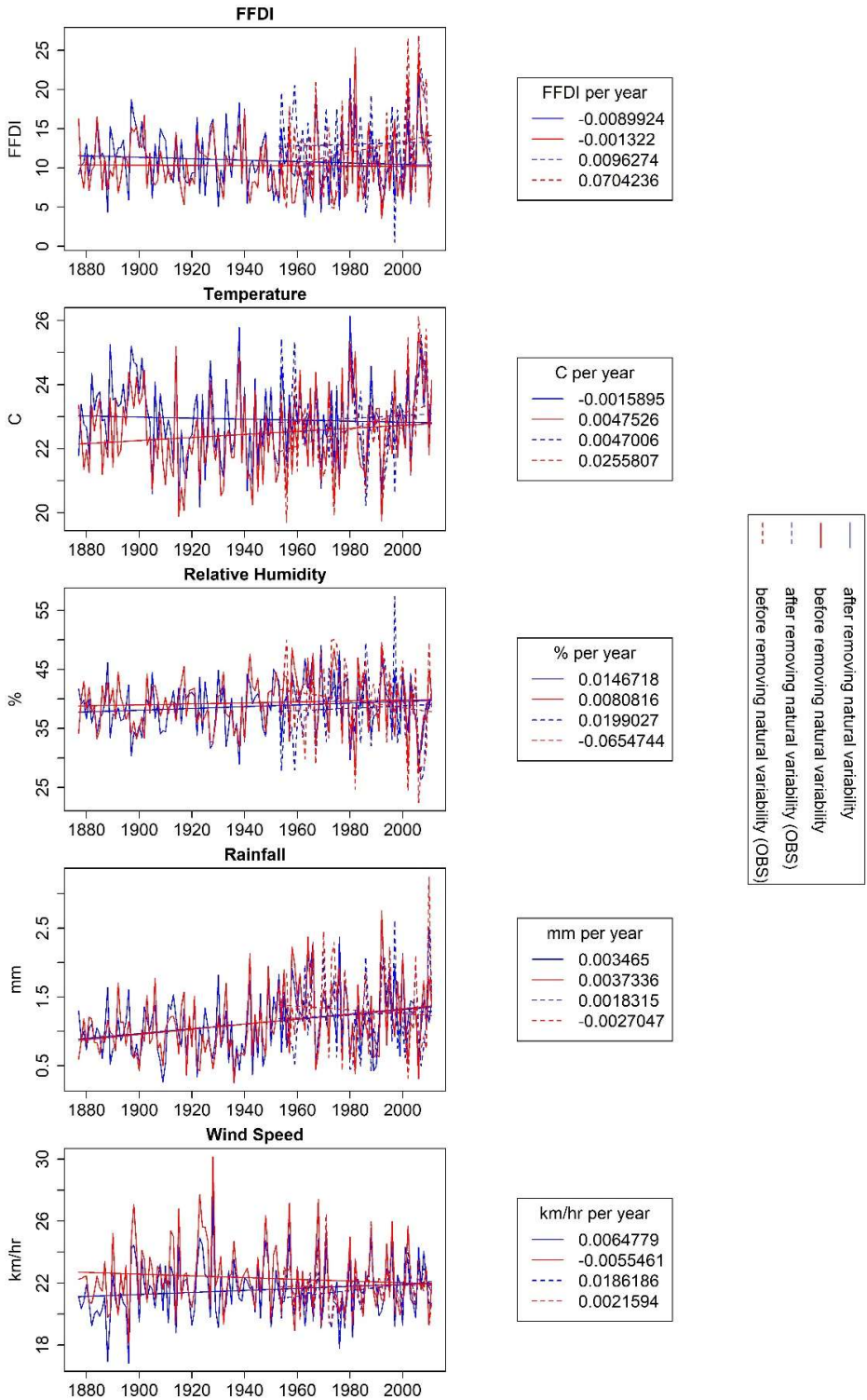

**Fig. 5.** Time series of mean seasonal FFDI, temperature, relative humidity, rainfall and wind speed for the austral spring season (SON) at Lockhart (146.72° E, 35.22° S) from 1877 to 2011. The red and blue dotted line indicates the observation-based values for the corresponding variable before and after removing the natural variabilities. The red and blue solid lines indicate the bias-corrected 20CRv2c values for the corresponding variable before and after removing the natural variabilities.

Similarly, for extreme FFDI days (Fig. 4), a strong significant upward trend is indicated across Australia during each of the four seasons, except for Murray Basin and the Southern Slopes during spring, where a statistically insignificant downward trend in extreme FFDI days was observed. There are still some notable differences in the spatial patterns of trend magnitudes compared to mean seasonal FFDI. For example, in contrast to the change in mean seasonal FFDI during the spring season, the Wet Tropics showed the strongest upward trends in extreme FFDI days (~0.06 year$^{-1}$), closely followed by the Rangelands (~0.05 year$^{-1}$) (Fig. 4d and Table 2). Significant positive trends were observed throughout Australia during the austral summer season, with a slightly higher trend magnitude over the Central slopes and East coast (Fig. 4a). Similarly, we again observe a statistically significant upward trend for the austral autumn and winter seasons across the entire Australian region (Fig. 4b, c). In the spring season, statistically significant upward trends are noted over most of Australia, except for NSW, Tasmania and Victoria (Fig. 4d). Results from the changes in extreme FFDI days (Fig. 4) between the two epochs are consistent with the spatial patterns in long-term trend magnitudes (Fig. 4) during each of the four seasons (Table 2). Both metrics show that fire weather conditions have worsened more during the recent period, as indicated by higher values for FFDI-based measures. A key result from Table 2 for this analysis of extreme values is that all of the significant trends are positive, with no significant negative trends, as was also the case for the results based on mean values from Table 1.

**Table 2.** As in Table 1 but for seasonal 90th percentile peak over threshold values of FFDI.

|  | DJF | MAM | JJA | SON |
| --- | --- | --- | --- | --- |
| Australia | **0.0529 (50.09%)** | **0.0677 (61.51%)** | **0.0709 (56.78%)** | **0.0469 (36.41%)** |
| Central Slopes | **0.0661 (66.75%)** | **0.0869 (81.71%)** | **0.0704 (63.16%)** | 0.0329 (10.49%) |
| East Coast | **0.0769 (74.95%)** | **0.0857 (67.84%)** | **0.0709 (75.79%)** | **0.0527 (39.32%)** |
| Murray Basin | **0.0466 (49.10%)** | **0.0476 (39.34%)** | **0.0499 (36.93%)** | 0.0006 (-3.27%) |
| Monsoonal North | **0.0541 (51.68%)** | **0.0624 (47.65%)** | **0.0553 (34.30%)** | **0.0505 (34.36%)** |
| Rangelands | **0.0528 (49.16%)** | **0.0706 (68.40%)** | **0.0802 (67.69%)** | **0.0535 (45.71%)** |
| Southern Slopes | **0.0334 (42.55%)** | **0.0522 (52.48%)** | **0.0512 (42.38%)** | **0.0203 (14.93%)** |
| Southern and South-Western Flatlands | **0.0377 (40.66%)** | **0.0492 (53.18%)** | **0.0637 (49.75%)** | **0.0324 (31.07%)** |
| Wet Tropics | **0.0440 (36.34%)** | **0.0499** (14.98%) | **0.0359 (17.43%)** | **0.0616 (42.37%)** |

Similarly, here also, relative humidity (Fig. 6) seems to be the most influential climatic variable, closely followed by temperature (Fig. 7) and rainfall (Supplementary Fig. 5). There is another important aspect to be noted here while analysing the spatio-temporal variations of extreme days for relative humidity. We have considered the number of days in each season that are below the 10th percentile to be extreme days, while for other variables, we considered the number of days each season above the 90th percentile to be extreme days.

This was done to better demonstrate the influence of dry days on extreme FFDI days. Likewise, we again saw a strong increasing trend across Australia for each of the four seasons except for the Murray Basin during the spring season. This pattern in extreme relative humidity days per year closely resembles that of trends in extreme FFDI days. Trends in extreme temperature days are generally consistent with the trends in extreme FFDI days for most regions across Australia. For example, we see a statistically significant increasing trend in extreme temperature days for all seasons, with exceptions for regions across southwestern Australia and South Australia (Fig. 7d), where we note slightly suppressed (but statistically insignificant) trends in extreme days for temperature. As for the trends in extreme days for wind speed (Supplementary Fig. 8), it has similarities in some regions with trends in seasonal extreme FFDI days (Fig. 4). For example, during the austral spring season, both extreme FFDI days and extreme wind speed days show a significant increasing trend over the Central slopes, East Coast, Monsoonal North and Rangelands (Fig. 4d and Supplementary Fig. 8d). But in the case of the summer season, we notice a spatial similarity only over the Monsoonal North, Central slopes and Southern slopes (Fig. 4a and Supplementary Fig. 8a).

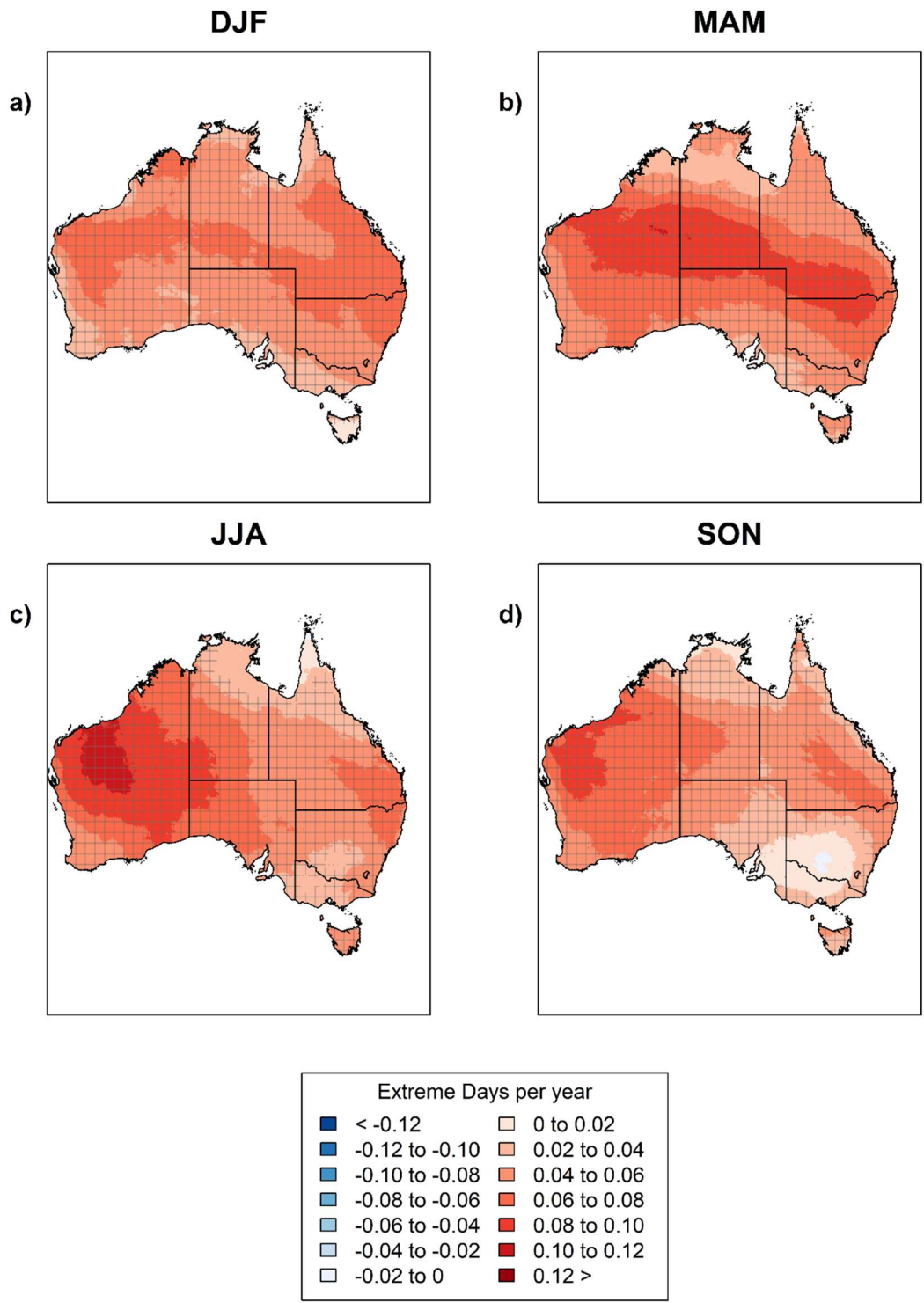

**Fig. 6.** As in Fig. 4 but for seasonal 10th percentile peak under threshold values of relative humidity.

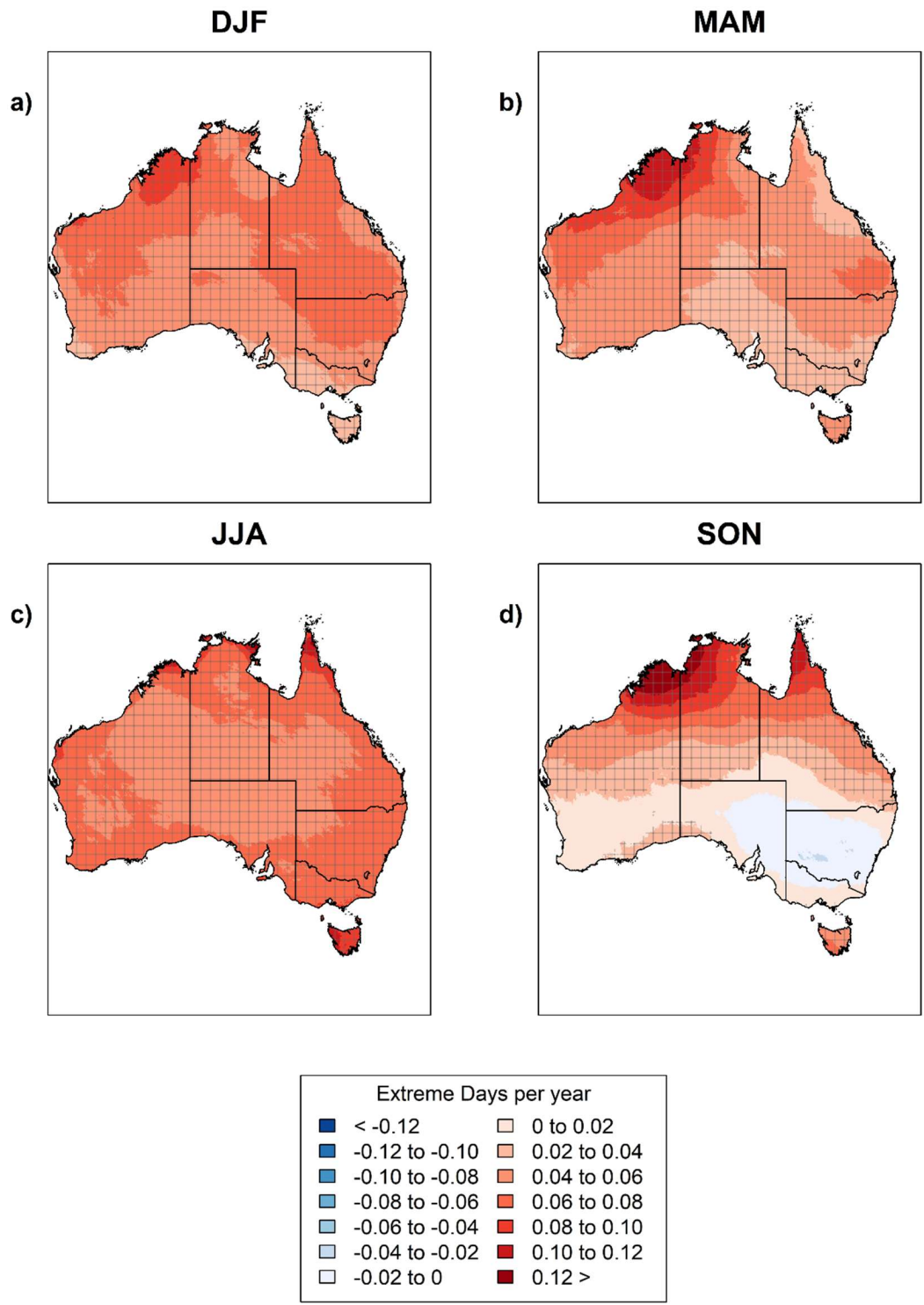

**Fig. 7.** As in Fig. 4 but for seasonal 90th percentile peak over threshold values of temperature.

Results were also examined based on using other time periods from within the full 20CR period of coverage, such as the second half of the study period used here. For example, the mean-state changes in seasonal FFDI values (Supplementary Fig. 9 and Table 3) and the changes in extreme FFDI days (Supplementary Fig. 10 and Table 4) between the two climatic epochs, 1945-1978 and 1978-2011, were calculated to understand the changes in Australian fire weather after the era of more rapid industrialization towards the later part of the 20$^{th}$ Century. Mean FFDI values show an increase across eastern Australia during Austral Summer (Supplementary Fig. 9a) and autumn (Supplementary Fig. 9b) and a statistically non-significant decrease over Western Australia. Austral Spring (Supplementary Fig. 9d) and winter (Supplementary Fig. 9c) also showed an increase in FFDI value but were lower in intensity when compared to Supplementary Fig. 1. Similarly, for extreme FFDI days (Supplementary Fig. 10), we again see an increase in extreme FFDI days in the second epoch (1978-2011) for most areas across Australia, but lower in intensity when compared to Supplementary Fig. 2. One explanation for this could be the missing FFDI data for the last decade, but it was not possible, as discussed in Data and Methodology (Dickman & McDonald, 2020; Filkov et al., 2020; Prelgauskas, 2016; Wilkinson et al., 2016), which could have altered the statistics presented here using data from 1945-2011.

## 4    Discussion

This study explores long-term spatio-temporal variations in fire weather conditions across Australia, including using data and methods complementary to previous studies. This has been achieved through the study of long-term climatological maps, seasonal changes and trends of FFDI. This work produced results broadly similar to prior studies (Canadell et al., 2021; Clarke et al., 2013; Dowdy, 2018; Dowdy & Pepler, 2018; Harris & Lucas, 2019), while noting one key point of difference for this study being its use of the long-term FFDI data back to 1876. In addition to using methods similar to previous studies, additional approaches were also applied, including a method to help account for natural variability for enhanced insight on the trends.

This is the first time for any region of the world, as far as the authors are aware, that long-term trends in fire weather have been examined that far back in time based on reanalysis data, in this case derived from bias-corrected 20CRv2c input climatic variables. This is complementary to previous analyses, such as long-term trend analysis based on Australian station observations[14,16], as well as noting some palaeontology studies that have used proxy methods for considering fire occurrence over time (Fletcher et al., 2021; Luhar et al., 2020).

Trends towards more dangerous fire weather conditions were found to be primarily driven by reductions in relative humidity and higher temperatures, noting that relative humidity is in part based on temperature (as well as atmospheric moisture content). Trends towards higher temperatures in Australia, as shown here, as well as globally, are attributable to anthropogenic climate change. For example, as detailed in (IPCC, 2023), land regions are warming rapidly, with oceans warming at a somewhat slower pace, which leads to general reductions in relative humidity over land. As such, the results are consistent with general expectations based on the climate change observed over the study period.

*Our key conclusions are:*

1. Positive trends in both mean and extreme FFDI days are evident through Australia for each of the four seasons, with all statistically significant trends being positive in sign for individual climate zones.
2. Humidity and temperature trends, attributable to human-caused climate change, are shown to be the main cause of the increase in dangerous weather conditions for fires.
3. The increased frequency of dangerous fire weather conditions is partly due to increased variability (including noting increased rainfall variability) as well as a trend in mean values, noting that this increase appears to be nonlinear over the long time period examined here, with acceleration in the more recent decades.
4. These findings build on previous studies, with the novel data and methods used adding confidence to the overall understanding of fire risk factors in a changing climate.

These results underline the role climate change played in exacerbating dangerous fire weather conditions across Australia, especially during the second half of the 20$^{th}$ Century, with more rapid industrialisation (e.g., post-1950s) than the earlier part of this study period, back to 1876. Improved knowledge of peak fire season over different climatic regions across Australia is vital for a number of fields, including disaster preparedness and management, ecological research and long-term government plans and policies.

**Data availability**

The 20th-century Version 2c (20CRv2c) data used in this study are available from the NERSC Science Gateway https://portal.nersc.gov/project/20C_Reanalysis/. The various datasets generated in our current study from 20CRv2c are not yet available on any pubic servers but are available from the corresponding author upon reasonable request.

**Conflicts of interest**

The authors declare that they have no conflicts of interest.

**Author contributions**

Dr. Soubhik Biswas, Dr. Savin S. Chand and Dr. Andrew J. Dowdy all contributed to the idea and design of the study. Soubhik Biswas prepared the material, collected the data, and conducted the analysis. Dr. Soubhik Biswas wrote the manuscript's initial draft, and all of the other authors offered feedback on earlier drafts. The final manuscript was read and approved by all authors.

## Acknowledgements

This research is funded by Henry Sutton PhD scholarship from Federation University. Support for the Twentieth Century Reanalysis Project version 2c dataset is provided by the U.S. Department of Energy, Office of Science Biological and Environmental Research (BER), and by the National Oceanic and Atmospheric Administration Climate Program Office.

Cameron, P. A., Mitra, B., Fitzgerald, M., Scheinkestel, C. D., Stripp, A., Batey, C., Niggemeyer, L., Truesdale, M., Holman, P., Mehra, R., Wasiak, J., & Cleland, H. (2009). Black Saturday: the immediate impact of the February 2009 bushfires in Victoria, Australia. *The Medical Journal of Australia*, *191*(1), 11-16. https://doi.org/10.5694/j.1326-5377.2009.tb02666.x

Canadell, J. G., Meyer, C. P., Cook, G. D., Dowdy, A., Briggs, P. R., Knauer, J., Pepler, A., & Haverd, V. (2021). Multi-decadal increase of forest burned area in Australia is linked to climate change. *Nature Communications*, *12*(1), 6921. https://doi.org/10.1038/s41467-021-27225-4

Christoff, P. (2023). *Fires Next Time: Understanding Australia's Black Summer*. Melbourne Univ. Publishing.

Clarke, H., Lucas, C., & Smith, P. (2013). Changes in Australian fire weather between 1973 and 2010. *International Journal of Climatology*, *33*(4), 931-944. https://doi.org/10.1002/joc.3480

Compo, G. P., Whitaker, J. S., Sardeshmukh, P. D., Matsui, N., Allan, R. J., Yin, X., Gleason, B. E., Vose, R. S., Rutledge, G., Bessemoulin, P., Brönnimann, S., Brunet, M., Crouthamel, R. I., Grant, A. N., Groisman, P. Y., Jones, P. D., Kruk, M. C., Kruger, A. C., Marshall, G. J.,…Worley, S. J. (2011). The Twentieth Century Reanalysis Project. *Quarterly Journal of the Royal Meteorological Society*, *137*(654), 1-28. https://doi.org/10.1002/qj.776

D. Ward, M., & S. Ahlquist, J. (2018). Theory and Properties of Maximum Likelihood Operators. In *Maximum Likelihood for Social Science: Strategies for Analysis* (pp. 21-42). Cambridge University Press.

Deb, P., Moradkhani, H., Abbaszadeh, P., Kiem, A. S., Engström, J., Keellings, D., & Sharma, A. (2020). Causes of the Widespread 2019–2020 Australian Bushfire Season. *Earth's Future*, *8*(11), e2020EF001671. https://doi.org/10.1029/2020EF001671

Dickman, C., & McDonald, T. (2020). Some personal reflections on the present and future of Australia's fauna in an increasingly fire-prone continent. *Ecological Management & Restoration*, *21*(2), 86-96. https://doi.org/10.1111/emr.12403

Dowdy, A. J. (2018). Climatological Variability of Fire Weather in Australia. *Journal of Applied Meteorology and Climatology*, *57*(2), 221-234. https://doi.org/10.1175/JAMC-D-17-0167.1

Dowdy, A. J., Mills, G. A., Finkele, K., & Groot, W. d. (2009). *Australian fire weather as represented by the McArthur Forest Fire Danger Index and the Canadian Forest Fire Weather Index* (CAWCR Technical Report ; 10, Issue), Bureau of Meteorology, Melbourne, Australia, ISBN: 9781921605185.

Dowdy, A. J., & Pepler, A. (2018). Pyroconvection Risk in Australia: Climatological Changes in Atmospheric Stability and Surface Fire Weather Conditions. *Geophysical Research Letters*, *45*(4), 2005-2013. https://doi.org/10.1002/2017gl076654

Dowdy, A. J., Ye, H., Pepler, A., Thatcher, M., Osbrough, S. L., Evans, J. P., Di Virgilio, G., & McCarthy, N. (2019). Future changes in extreme weather and pyroconvection risk factors for Australian wildfires. *Scientific Reports*, *9*(1), 10073. https://doi.org/10.1038/s41598-019-46362-x

Filkov, A. I., Ngo, T., Matthews, S., Telfer, S., & Penman, T. D. (2020). Impact of Australia's catastrophic 2019/20 bushfire season on communities and environment. Retrospective analysis and current trends. *Journal of Safety Science and Resilience*, *1*(1), 44-56. https://doi.org/10.1016/j.jnlssr.2020.06.009
21

# Supplementary information

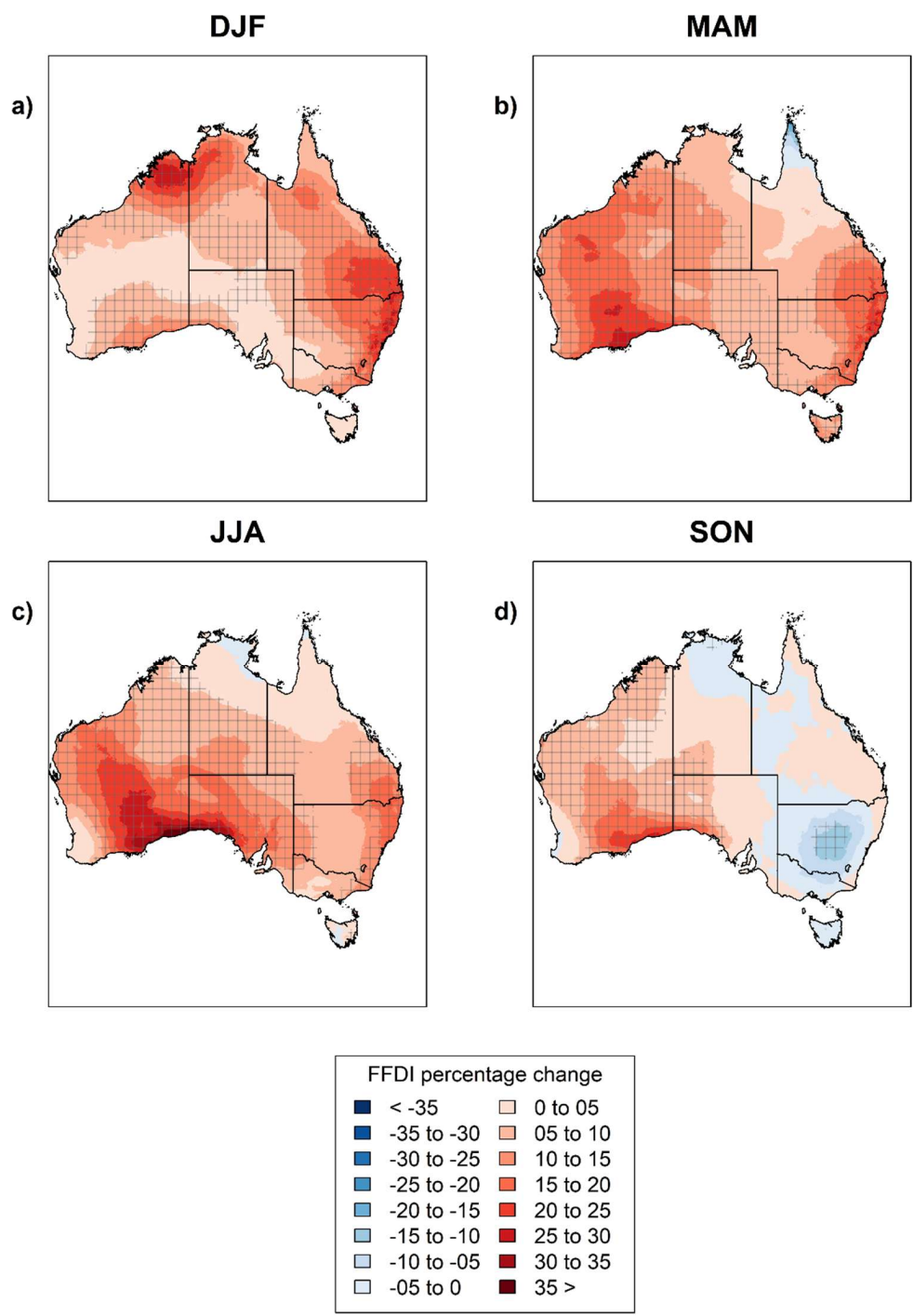

**Supplementary Fig. 1.** Percentage change in seasonal mean FFDI values between 1876 and 1945 and 1945 and 2011. This is shown by the percentage change in FFDI values between 1876 and 1945 and 1945 and 2011 for each of the four seasons. Statistically significant regions are shown as overlapping grey grids.

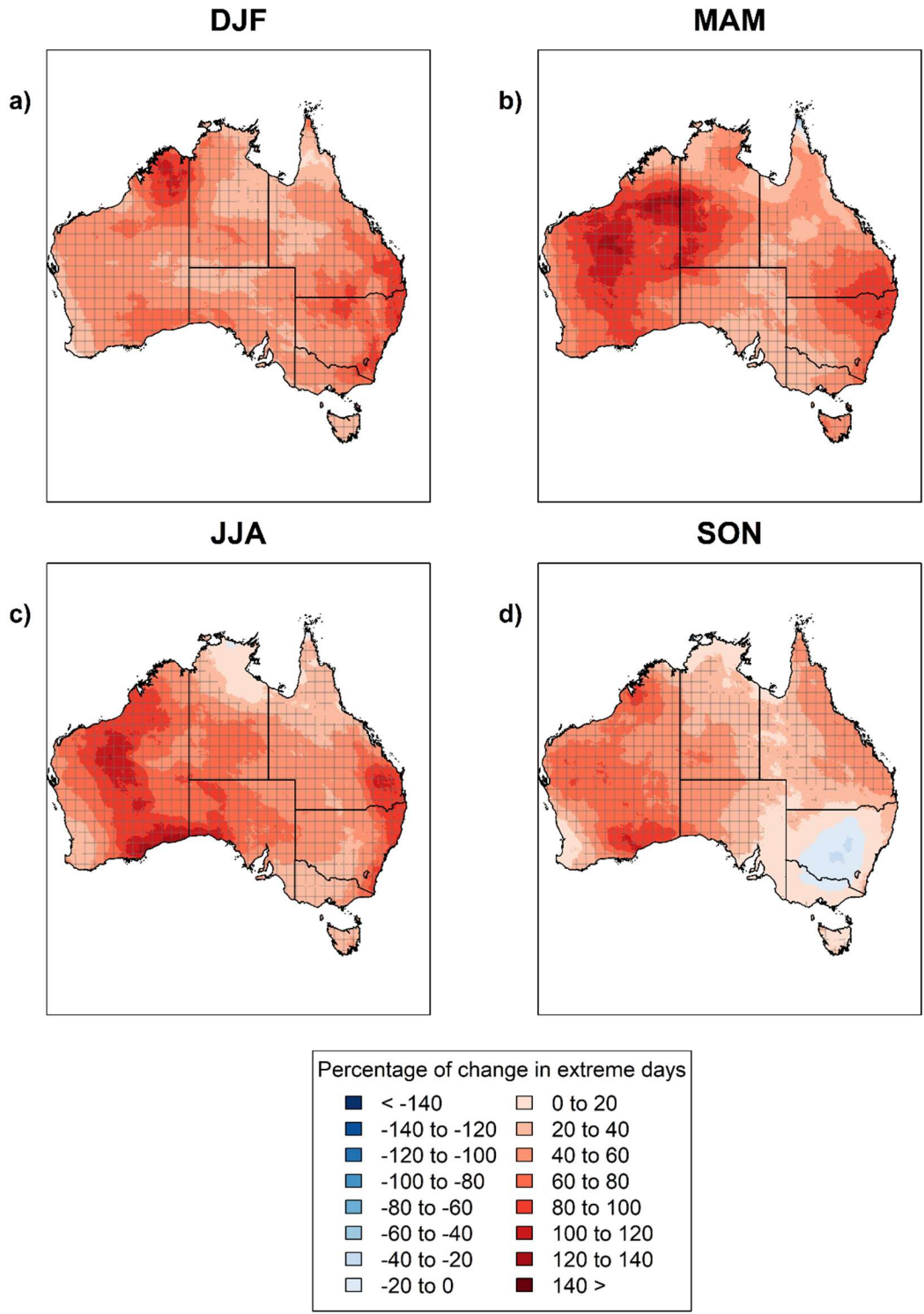

**Supplementary Fig. 2.** As in Fig 1, but for 90th percentile peak over threshold value of FFDI (i.e., the number of days per season that the FFDI is above its 90th percentile).

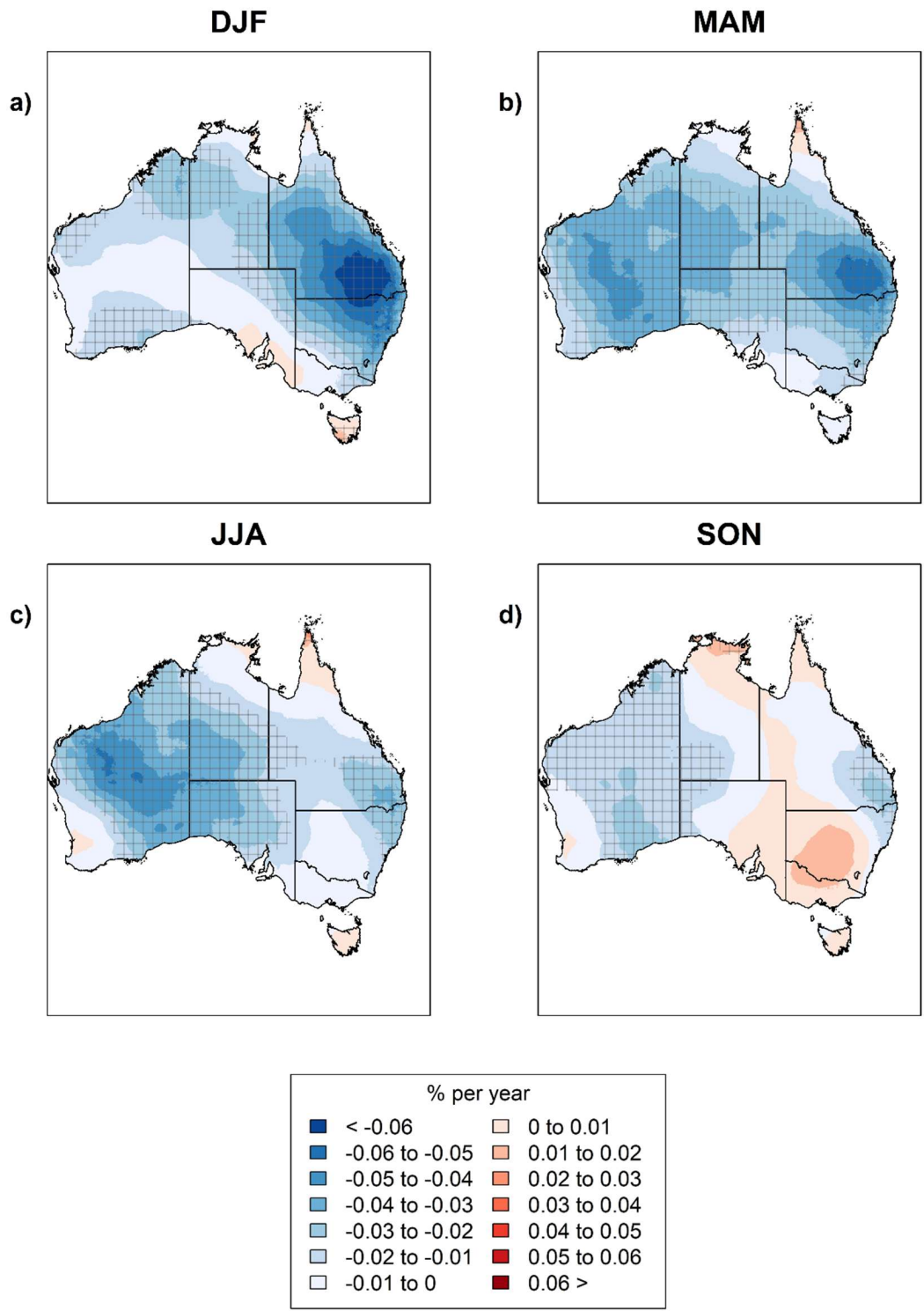

**Supplementary Fig. 3.** Mean seasonal trend in relative humidity at each grid point for the study period 1876 – 2011. Statistically significant regions are shown as overlapping grey grids.

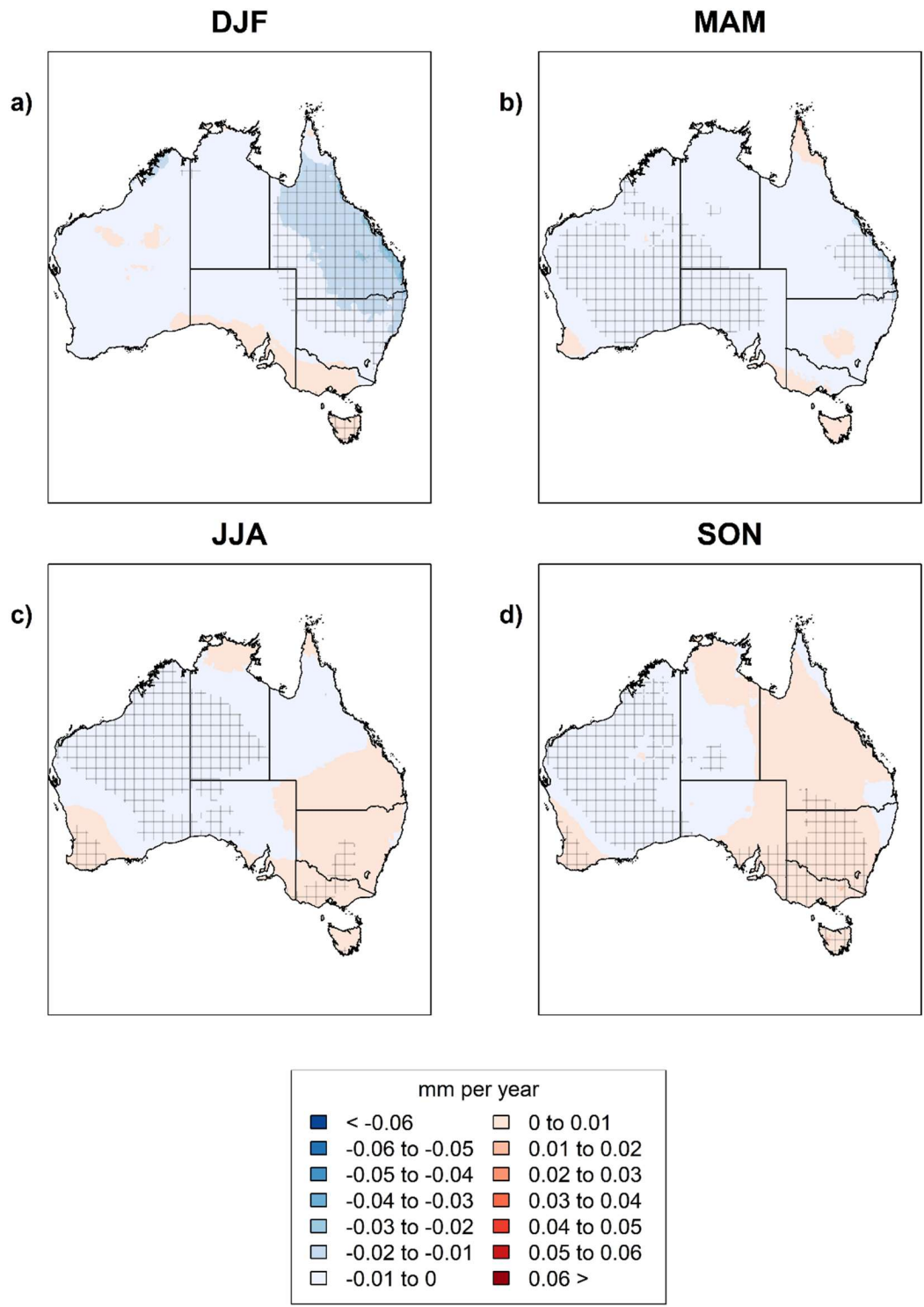

**Supplementary Fig. 4.** Mean seasonal trend in rainfall at each grid point for the study period 1876 – 2011. Statistically significant regions are shown as overlapping grey grids.

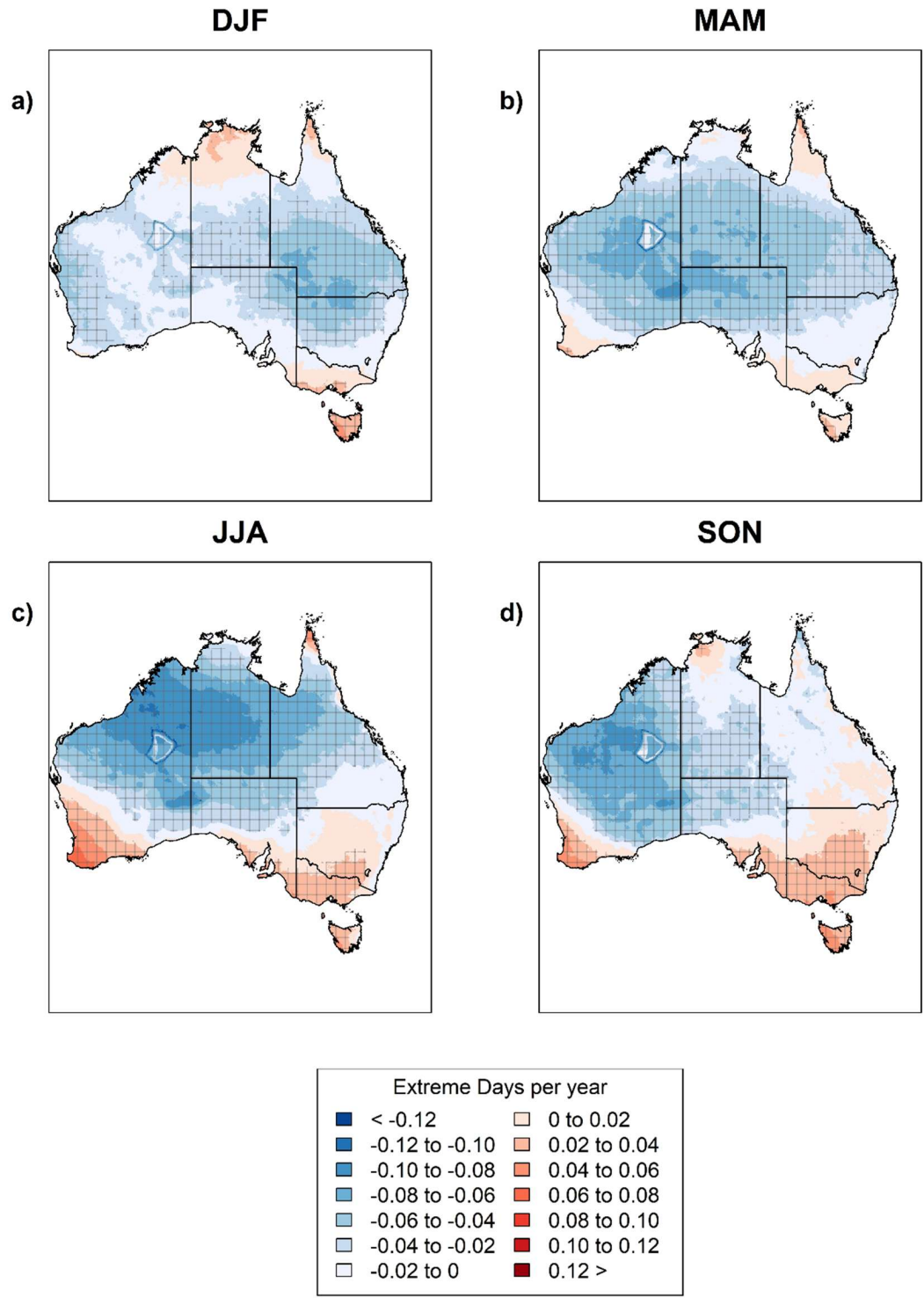

**Supplementary Fig. 5.** As in Supplementary Fig. 4 but for seasonal 90th percentile peak over threshold values of rainfall.

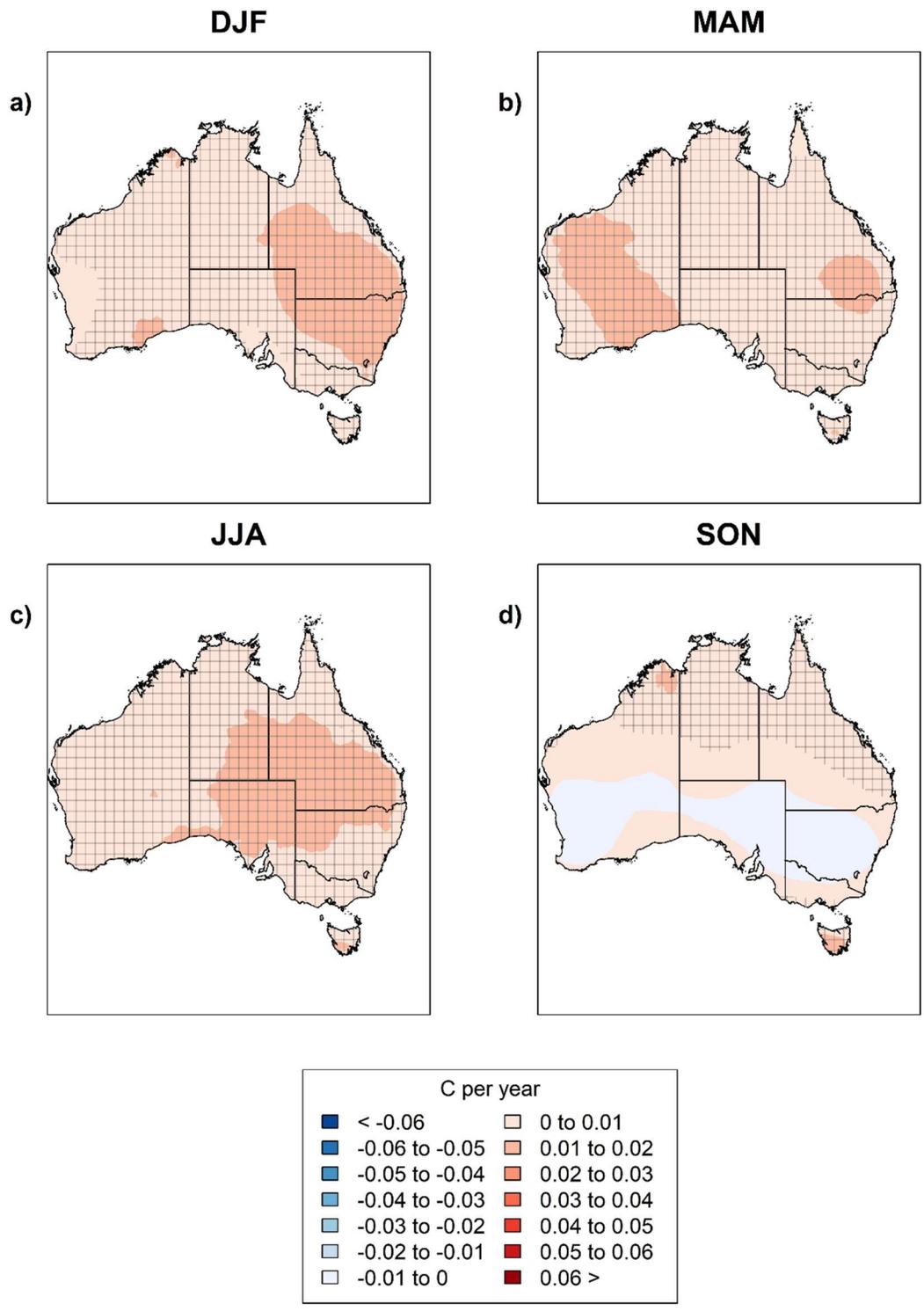

**Supplementary Fig. 6.** Mean seasonal trend in temperature at each grid point for the study period 1876 – 2011. Statistically significant regions are shown as overlapping grey grids.

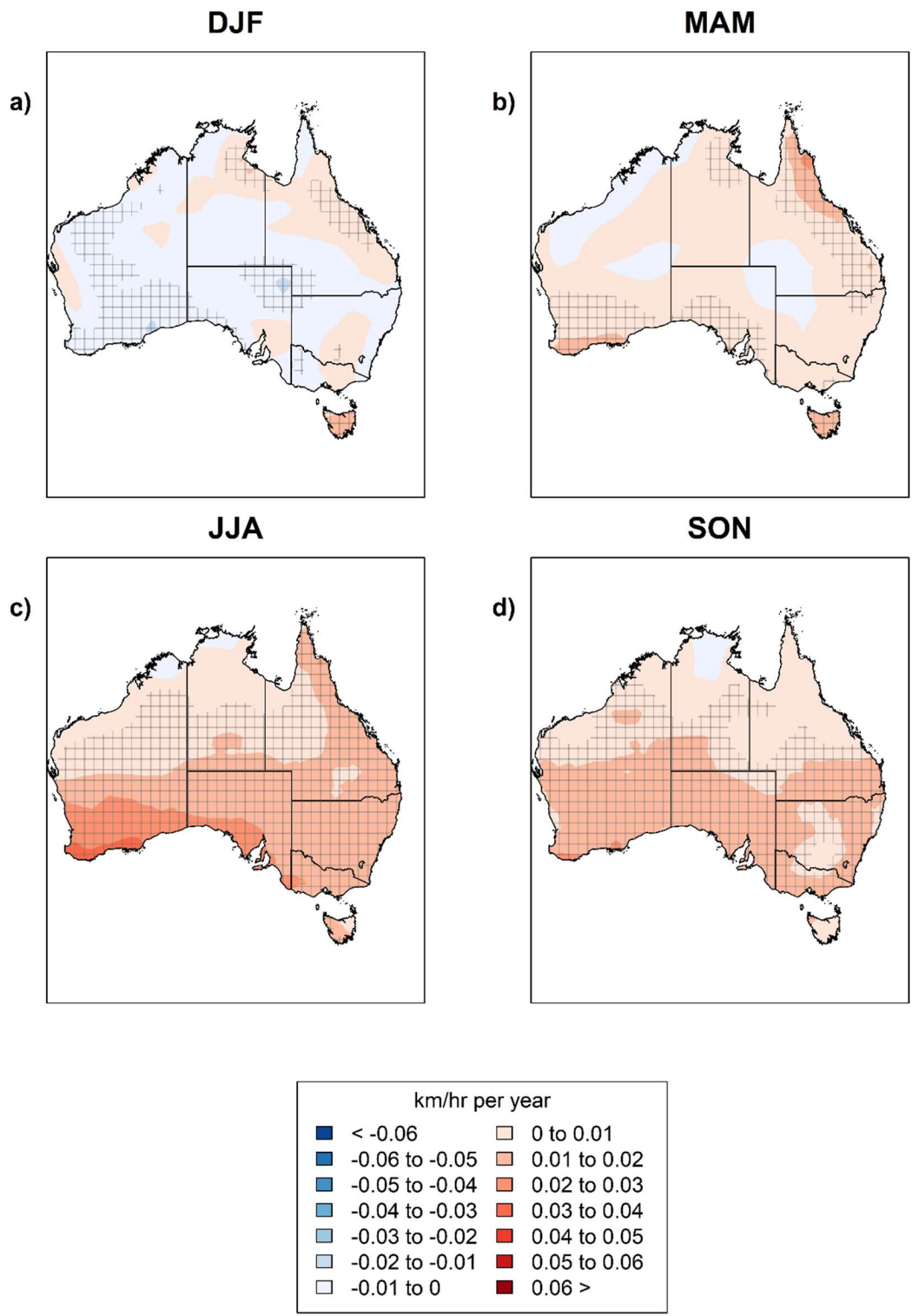

**Supplementary Fig. 7.** Mean seasonal trend in wind speed at each grid point for the study period 1876 – 2011. Statistically significant regions are shown as overlapping grey grids.

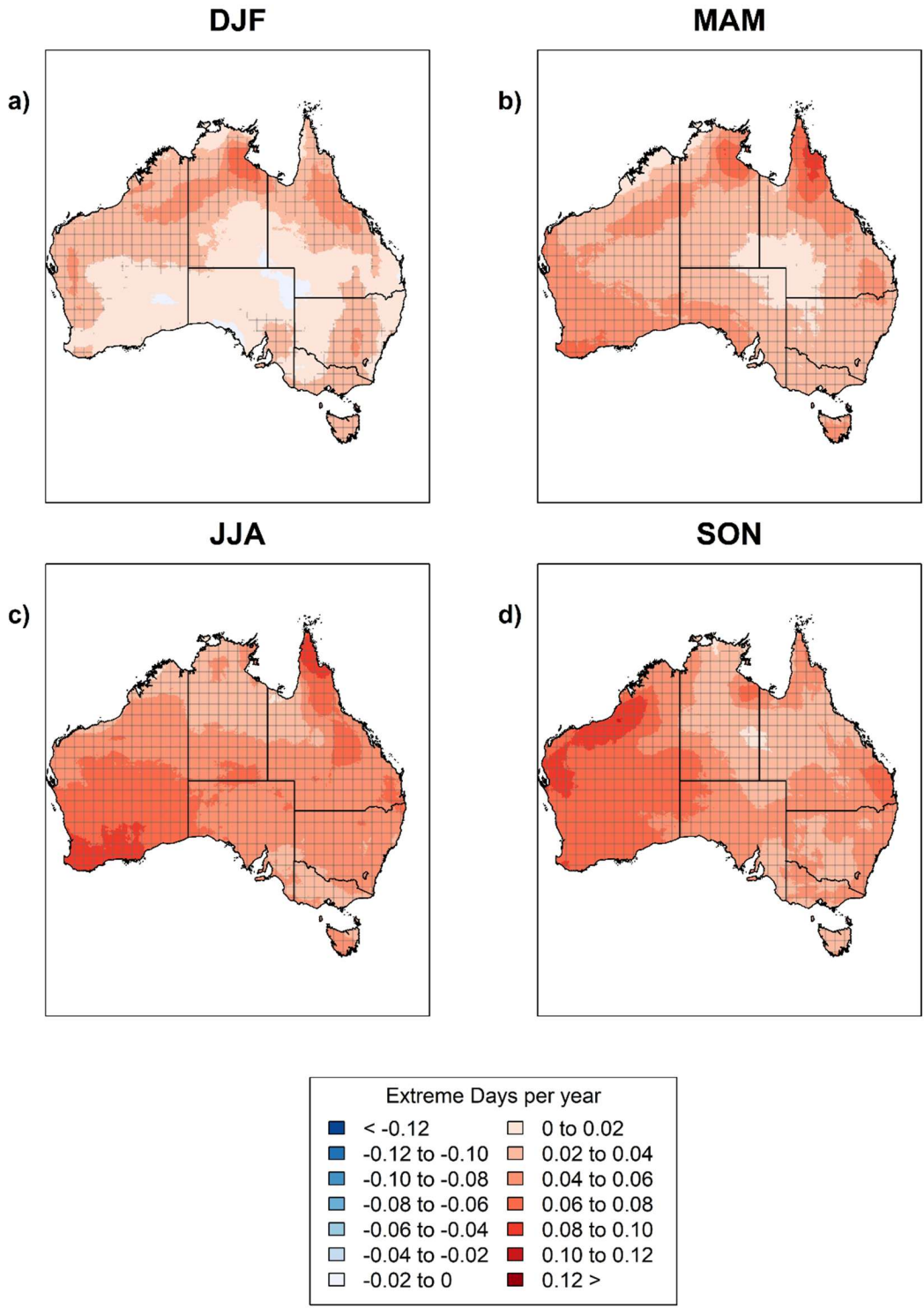

**Supplementary Fig. 8.** As in Supplementary Fig. 7 but for seasonal 90th percentile peak over threshold values of wind speed.

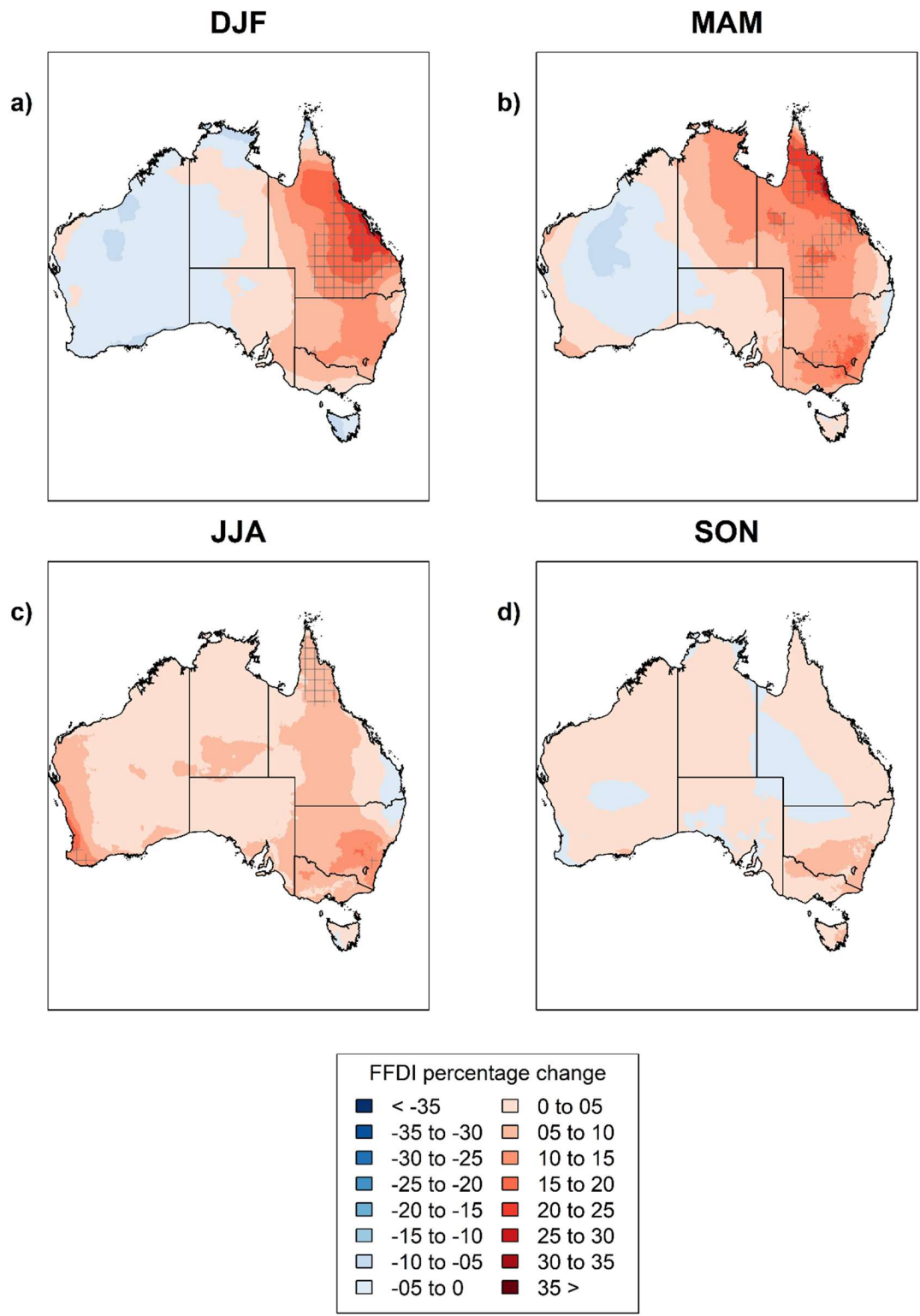

**Supplementary Fig. 9**. Percentage change in seasonal mean FFDI values between 1945 and 1978 and 1978 and 2011. This is shown by the percentage change in FFDI values between 1945 and 1978 and 1978 and 2011 for each of the four seasons. Statistically significant regions are shown as overlapping grey grids.

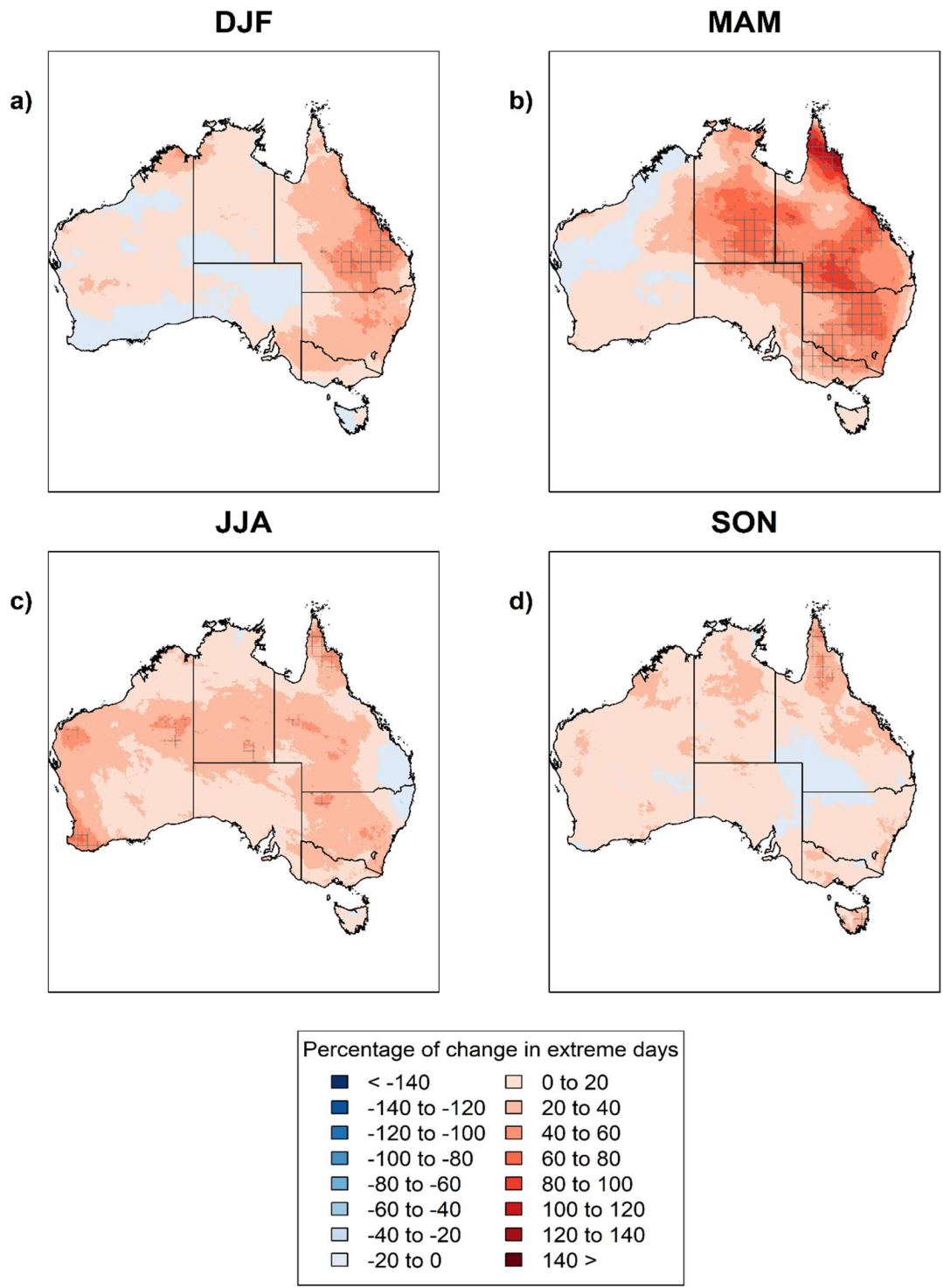

**Supplementary Fig. 10.** Percentage change in seasonal 90th percentile peak over threshold values of FFDI (i.e., the number of days per season that the FFDI is above its 90th percentile) between 1945 and 1978 and 1978 and 2011. This is shown by the percentage change in the number of FFDI extreme days between 1945 and 1978 and 1978 and 2011 for each of the four seasons. Statistically significant regions are shown as overlapping grey grids.